# A Study on Spectral Efficiency of Physical Layer over Cognitive Radio

Fotis Foukalas[1], Lazaros Merakos[1]

[1]*University of Athens, Department of Informatics and Telecommunications, Ilisia Athens, 15784, Greece*
*Tel: +302107275323, Fax: +302107275601,*
*Email: foukalas@di.uoa.gr, merakos@di.uoa.gr*

**Abstract:** Despite the conclusive potential of cognitive radio for provisioning the dynamic and flexible spectrum/channel allocation, the research community should study the performance gain of physical layer over such a radio with cognition capabilities. To this end, several mechanisms of physical layers such as adaptive modulation, multiple-input multiple output antennas; channel coding and/or combination of them should be studied. These studies should be accomplished in terms of spectral efficiency. Therefore, the gain of cognitive radio in wireless networks available into the market will be identified practically. Another issue under consideration should be the performance evaluation of cognitive radio assuming a cross-layer combination between the cognitive physical and the upper layers. To this direction, this paper presents a study on spectral efficiency at the physical layer with cognitive capabilities. In sequel, we study a cross-layer combination of physical layer with upper layers in the same cognitive context. The performance gain of cognitive radio in such a physical layer is realized practically as well as a few cross-layer design issues have been raised.

## 1. Introduction

The Cognitive Radio (CR) concept brought the idea to exploit the spectrum holes which result from the underutilization of the electromagnetic spectrum in wireless communications. This fact is corroborated by the Spectrum Policy Task Force of the Federal Communications Commission (FCC) which ascertains that the legacy regulation on spectrum availability begets snags in potential spectrum access by users. More precisely, it was identified that spectrum bands seems to be unoccupied most of the time and some of them seems to be occupied partially by the primary (or licensed) users [1].

Primary users are assigned with a range of frequency bands but however they do not use those one hundred percent in time or location. As a consequence, a particular communication system which serves its primary users at a specific geographic location may present spectrum holes which could be exploited by using a cognitive radio approach which can identify the possible holes in spectrum bands. In this way, secondary users that were not being served by the system can access and exploit the spectrum holes thus improving spectrum utilization.

The utilization of a cognitive radio approach requires the specification of the techniques by which primary users and secondary users will be served. Among the cognitive tasks of such a system the most prominent ones should be the reliable sense of spectrum range, the detection of spectrum holes, the estimation of mutual interference between primary and secondary users and the control of their transmission power as well. To this end, we assume a spectrum pooling system with cognition capabilities. Such a system is capable to reliably sense the spectrum range. Moreover, considering that the frequency carrier is divided into time slots then the mobile users are being served in a multi-band context. In such a context, the task of detecting holes in spectrum bands could be performed for instance by a 'listen-before-talk' strategy. This strategy is able to detect the sub-bands (i.e. the time slots) which are currently available for assignment to secondary users [2].

On the other hand, spectrum pooling is able to gather available spectrum resources from other unlicensed (e.g. military radio systems) or licensed systems, which serve the same geographical area, and make them available to secondary users as long as it can sense the spectrum and detect the spectrum holes [3]. Due to the concurrent spectrum/channel allocation from the system, mutual interference may show up between the primary (licensed) and secondary (unlicensed) users. In addition to all these tasks, the cognitive radio channel allocation can rely on a power control policy with average optimum transmit power constraint in order to achieve the maximum instantaneous system capacity, in other words spectral efficiency.

By assuming such a cognitive radio system we should be able to assess the performance of physical layer. Generally speaking, the physical layer consists of the following mechanisms such as the adaptive modulation, the multiple-input multiple output antennas, the channel coding and the multiple access techniques. Assuming also the upper layers (i.e. data link and application layers), other mechanisms such as automatic-repeat request could be involved in this assessment. To this direction, we have chosen to study the performance of adaptive modulation over such a radio environment which represents the cognitive physical layer in our system model. In sequel, we implement a cross-layer design on such a cognitive physical layer. The cross-layer design involves constraints from the application layer which is being transformed in constraints imposed by the data link layer. In order to make an objective performance evaluation, we should study the cognitive radio system behaviour over a particular fading propagation environment. For this reason, before focusing on the aforementioned study, we provide the performance formulation for the capacity of a cognitive radio system over a flat fading channel with Rayleigh coefficients.

The rest of this paper is organized as follows. Section 2 presents the channel allocation with optimum transmit power constraint employed by the cognitive radio system. Section 3 gives the performance analysis of cognitive radio in fading environment. Section 4 describes the performance of adaptive modulation scheme over the aforementioned cognitive radio under the fading phenomena. Section 5 investigates the performance behaviour of cognitive physical layer under cross-layer design constraints which imposed from the upper layers. The paper is concluded with section 6 as the future work on performance studies of cognitive physical layer.

## 2. Channel Allocation in Cognitive Radio

In spectrum pooling system, the spectrum or channel allocation gives priority to primary or licensed users. In sequel, the secondary users are assigned the detected spectrum holes. These holes are assumed voids in sub-band range. Secondary users fill these voids as long as they achieve the desirable transmit power level. In other words, a channel is allocated to each secondary user based on optimum transmit power constraints. These power constraints are expressed by a threshold known as a channel gain.

More specific, the spectrum is divided into $N$ sub-bands and each user $l$ transmits when the channel gain is up a predefined threshold. The system transmission is considered as wide-band. In this case, the number of sub-bands $N$ of the cognitive radio channel extends to infinity ($N \rightarrow \infty$). The channel is assumed with fading components that is varying slow in time. Thus, the receiver is able to sense and track the channel fluctuations. These fluctuations are the aforementioned channel gains $h_l$ and they are assumed over a block fading length. This means that an instantaneous channel gain keeps its value constant throughout the period of processing a specific block. Figure 1 shows the channel allocation when eight sub-bands are considered. Primary users are initially served by the system and in sequel the secondary users is prioritized, in order to fill the already sensed holes in spectrum, according to the power control policy.

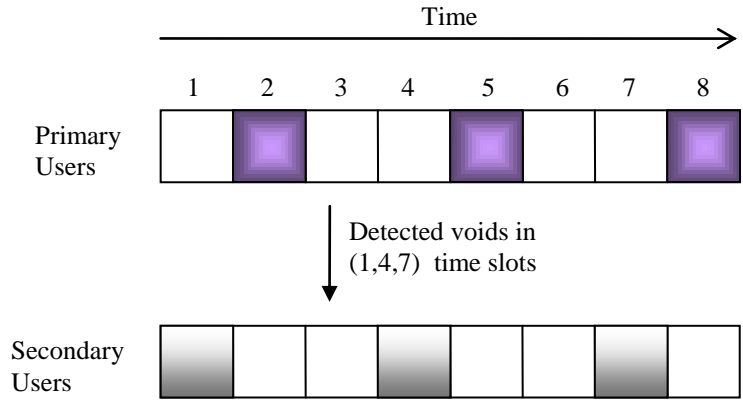

*Figure 1: The channel allocation in cognitive radio system*

Under these assumptions, the instantaneous capacity of user $l$ in bits/sec/Hz of the cognitive radio system is obtained by the following equation:

$$C_{l,\infty} = \int_0^\infty \left( \frac{1}{\gamma_0} - \frac{P_l(t)t}{N_0} \right) \cdot e^{-t} dt \quad (1)$$

where $l \in [1, L]$ is the number of users being served by the cognitive radio system with N sub-bands with $N \to \infty$. However, a channel is allocated to each user when the maximum transmission rate potentially can be achieved. Henceforth, the transmit power for each user $P_l$ is subject to the average constraint:

$$\int_0^\infty (\frac{1}{\gamma_0} - \frac{N_0}{t}) \cdot e^{-t} dt = 1 \quad (2)$$

This technique gives the optimal transmit power and rate adaptation and is known as water-filling. Based on equations (1) and (2), the cut-off SNR (signal-to-noise ratio) level is obtained that is equal to $\gamma_0 N_0$. The cut-off level in SNR indicates the minimum value of fade depth below which no signal is transmitted. In other words, it is the value of the channel gain above which a secondary user can transmit on the corresponding sub-band. We have to mention that the channel model in both equations (1) and (2) is specified by a general probability distribution function $e^{-t}$ [5]. In the next section, we present the appropriate changes on performance analysis of the aforementioned cognitive radio system in order to obtain the performance formulation of cognitive radio in fading environment.

## 3. Cognitive Radio in Fading Environment

It is evident that a cognitive radio system underlies in wireless propagation models, which despite the additive noise, they encompass fading phenomena as well. Moreover, since we study the performance of adaptive modulation, it is required the study of cognitive radio over fading channel. It is known that the adaptive modulation has been studied over Nakagami fading channels and thus we assume the case of Rayleigh model that is the case of m=1 in Nakagami channel model [4].

Nevertheless, taking into account the power control which the cognitive radio employs, we should study the spectral efficiency over Rayleigh fading channel given an average transmit power constraint which should be retained during the transmission. This is the optimal simultaneous power and rate adaptation case of study on fading channels [6]. Afterwards, assuming the Rayleigh probability density function for the SNR level denoted as $\gamma$, $p_\gamma(\gamma) = e^{-\gamma/\bar{\gamma}}/\bar{\gamma}$, where $\bar{\gamma}$ is the average received SNR, we obtain the optimal capacity of user 1 (i.e. the primary user) in cognitive radio system over Rayleigh fading channel in bits/sec/Hz as follows:

$$\langle C_{1,\infty} \rangle_{Rayleigh} = \log_2(e) \left( \frac{e^{-\gamma_0/\bar{\gamma}}}{\gamma_0/\bar{\gamma}} - \bar{\gamma} \right) \quad (3)$$

Having obtained the capacity of primary user, we should obtain the cut-off level which indicates the transmission of all users. In case of Rayleigh fading channels, the cut-off level is equal to $\gamma_0$ values and it is calculated by the following imposed average power constraint

$$\int_0^\infty \left( \frac{1}{\gamma_0} - \frac{1}{\gamma} \right) \cdot p_\gamma(\gamma) d\gamma = 1 \quad (4)$$

Afterwards, we are ready to define the spectral efficiency of cognitive radio over Rayleigh channel. In cognitive radio systems, the primary users are first being served by the channel allocation scheme, as described above. In this case, the spectral efficiency is equal to expression (3). However, when secondary users are appeared that need to be allocated with channels, then the spectral efficiency for each user is the equation (3) multiplied by the band factor gain. The band factor gain $\Delta_\infty$ of the cognitive radio system is defined as the band sensed void from user $l$ to user $l+1$ over the total bandwidth $W$ that is obtained as follows:

$$\Delta_\infty = 1 - \exp\left(-\frac{\gamma_0}{\bar{\gamma}}\right) \quad (5)$$

Therefore, the spectral efficiency of user $l$ is defined as follows

$$C_{l,\infty} = \Delta_\infty \cdot C_{l,\infty} \quad (6)$$

where $l = 1,...,L$. The overall system performance, i.e. the sum spectral efficiency for a system with $L$ users, is calculated by the approximation:

$$Se_{sum,\infty} = \frac{1 - \Delta_\infty^L}{1 - \Delta_\infty} \cdot C_{1,\infty} \quad (7)$$

The numerical results depicted in Figure 2 corroborate the above analysis in conjunction with Rayleigh analysis presented in [6]. Particularly, it illustrates the spectral efficiency of optimal power and rate adaptation using the closed-form expression (3) with red diamonds. Moreover, we draw the sum spectral efficiency of cognitive radio assuming Rayleigh fading channel. This is done assuming the closed-form expression (7) with $\gamma_0/\bar{\gamma}$ channel gains as derived the power constraint (i.e. equation (4)). Considering $L=5$ users for the cognitive radio system, the black line with circles represents the overall system performance (i.e. the sum spectral efficiency) of the cognitive radio over Rayleign channel. In order to corroborate the correctness of our results, we draw the system performance when one user is considered with green line. As expected, the cognitive radio assumption yields an increase in capacity that has a considerable value in low average SNR regions that is ranged between 0.2-0.3 bits/sec/Hz. Afterwards, we are ready to proceed towards the performance analysis of adaptive modulation over cognitive radio.

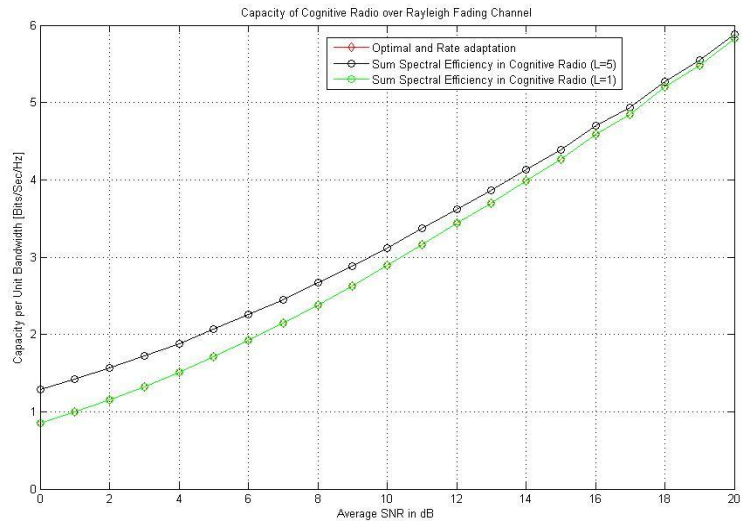

*Figure 2: The spectral efficiency of cognitive radio in Rayleigh fading*

## 4. Adaptive Modulation over Cognitive Radio

Adaptive modulation is the well-known scheme in which the transceiver is able to adapt its rate and/or power according to the channel variation that can be estimated at the receiver side. Adaptive modulation is well considered for multi-path fading channels [4][7]. Although, it was introduced first in [7], studying variable rate and variable power (VRVP) modulations, in sequel it was considered assuming variable-rate and constant-power (VRCP) for each transmission attempt [4]. It is evident that the VRVP instead of VRCP uses power adaptation that is relied on optimal power control policy.

As described above, the achievable performance of cognitive radio system depends on the band factor gain. This factor implies the cognition capabilities of each user (i.e. receiver) to sense the holes in the spectrum range which in sequel are fulfilled by the corresponding channel allocation technique. We have already mentioned that the channel allocation employs a power control policy which depends on the cut-off level in frequency variation. The obvious inference is that we use the VRVP case for studying the performance of adaptive modulation over cognitive radio which is based on optimal power control policy as well.

To this end, we should rely on the performance of VRVP case of adaptive modulation and particularly on its spectral efficiency [4]. The VRVP scheme employs MQAM constellations denoted as a set $\{M_j : j = 0,1,...,N\}$ which can be chosen according to the fade level $\gamma$ during the symbol period. In this set, the choice $M_0$ means no data transmission while $\{M_1,...,M_N\}$ denotes the constellation order in accordance with the corresponding fading regions denoted as $\{\gamma_1,...,\gamma_{N+1}\}$.

Thus, when the fading level is set in the $jth$ region then the constellation $M_j$ is chosen and in consequence the current data rate of the adaptive system is $\log_2 M_j$. However, since the transmit power $S(\gamma)$ should also be adapted, in order to retain the average power constraint $\bar{S}$, the received SNR is equal to $\gamma \cdot S(\gamma)/\bar{S}$. Henceforth, the scheme should be able to decide in which rate and transmit power it will operate in the next period. Both rate and power adaptation process is related to the required BER value (target BER). This target value must be offered by the adaptive modulation technique at the physical layer.

Afterwards, the aim of such a system, i.e. that involves power control policy, is to maximize the spectral efficiency subject to the power constraint

$$\frac{S(\gamma)}{\overline{S}} = \begin{cases} \frac{1}{\gamma_0} - \frac{1}{\gamma K}, & \gamma \geq \gamma_0 / K \\ 0, & \gamma \prec \gamma_0 / K \end{cases} \quad (8)$$

It is obvious that in this case the cut-off level in fading channel is equal to $\gamma_0 / K$ where K is in relation with the target BER ($K = -1.5/\ln(5BER)$). Therefore, denoted the $\gamma_0 / K$ as $\gamma_K$ then the spectral efficiency is maximized when the data rates are fallen in the fading region with optimum power allocation expressed as $\gamma_K / \overline{\gamma}$. In other words, the channel gain is equal to $\gamma_K / \overline{\gamma}$.

Afterwards, this dependency between the band factor gain and the optimum power control policy in order to maximize the spectral efficiency of VRVP adaptive modulation, is expressed as follows

$$\langle \Delta_\infty \rangle_{VRVP} = 1 - \exp(-\gamma_K / \overline{\gamma}) \quad (9)$$

Taking into account the sum spectral efficiency of cognitive radio, we derive the sum spectral efficiency of adaptive modulation over cognitive radio as follows

$$\langle Se_{sum,\infty} \rangle_{VRVP} = \frac{1 - \langle \Delta_\infty \rangle_{VRVP}^L}{1 - \langle \Delta_\infty \rangle_{VRVP}} \cdot \left( \frac{R}{B} \right) \quad (10)$$

where $R/B$ denotes the spectral efficiency of VRVP scheme which follows the aforementioned power control policy expressed by equation (8).

Figure 3 shows the results obtained by (10). We depict also the performance results of VRVP adaptive modulation without cognition capabilities that can be also obtained by using equation (10) with L=1 user. The regions denote the number of constellation that the system employs i.e. the number of deployed M-QAMs at the physical layer. The performance gain of cognitive radio over adaptive modulation with variable-rate and variable-power adaptation policy can be realized by Figure 3.

The lesson to be learned is first that the performance gain is increased as the fading regions are increased either. Moreover, when 5 regions are considered then the performance gain in terms of bps/symbol is approximately 0,4 in high average SNR regions. In case of 4 regions, this gain seems to be close to 0.2 bps/Hz while in 3 regions case it is 0.1 bps/Hz. Therefore, it could be argued that the performance gain of adaptive modulation over cognitive radio is duplicated when the fading regions are increased by one. Finally, one should note that this gain is not extremely remarkable at the low average SNR regions; indeed one can say that this gain is not discriminated at all.

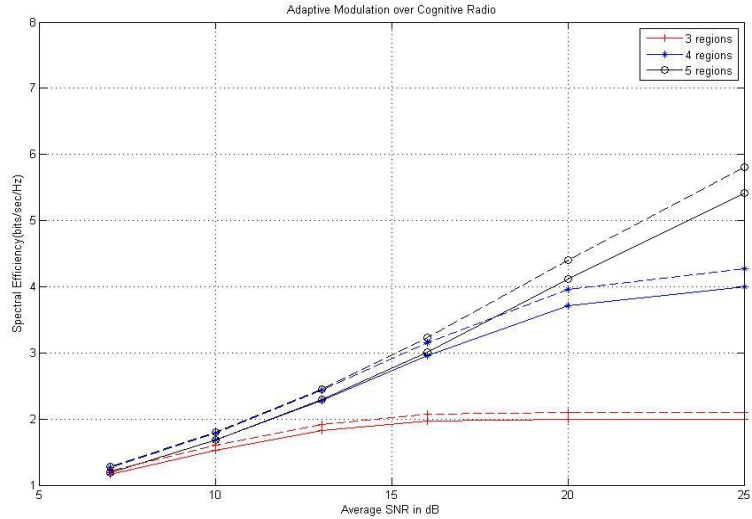

*Figure 3: The spectral efficiency of adaptive modulation over cognitive radio*

## 5. Cross-layer combination over Cognitive Radio

Since several approaches mostly in wireless communications involve the notion of cross-layer design [8], we suppose that such a study in cognitive radio is an interesting research topic in the area of cognitive radio. To this direction, we adopt the cross-layer design approach that combines adaptive modulation and hybrid automatic-repeat request at the physical and data link layer respectively [9]. This cross-layer approach in particular, is intended to co-design the adaptive modulation scheme at the physical layer taking into consideration a specific threshold in erroneous packets. The tolerable errors in packets are imposed by the quality of service that should be retained at application layers. Afterwards, the whole system is constrained by the number of packet retransmissions in order to keep the aforementioned quality criterion. This could be said that is an up to bottom cross-layer design approach [10].

More specific, the performance at the physical layer is equal to the spectral efficiency of adaptive modulation, which is the one without involving the aforementioned cross-layer design constraint, divided by the average number of retransmission at the data link layer. Thus, the average spectral efficiency is obtained as

$$\overline{S}_{e,cld} = \frac{\overline{S}_{e,physical}}{\overline{N}(p, N_t^{\max})} \quad (11)$$

where the term $\overline{S}_{e,physical}$ denotes the spectral efficiency at the physical layer and this is is actually the spectral efficiency of AMC-only scheme. On the other hand, the term $\overline{N}(p, N_t^{\max})$ expresses the average number of retransmission at the data link layer and it is related to packet error rate $p$ and the maximum number of transmission at the data link layer. For more details on denominator's definition, readers can refer to [9].

We suppose that the $\overline{S}_{e,physical}$ of the system is the spectral efficiency of the VRVP scheme. Having defined this spectral efficiency at the physical layer, we need to calculate the average number of retransmission at the data link layer under the aforementioned constraint. To this end, we consider 5 regions for the VRVP scheme with constellation ordering represented by the set of BPSK, QPSK, 16-QAM and 64-QAM. The numerical results derived by equation (11) is relied on the average PER at the data link layer which in sequel depends on the switching thresholds of the adaptive scheme. In our case, the performance results do not conclude in normal values if the appropriate changes are not provided. These changes are related to lower switching thresholds that the optimum VRVP reveals. This results in full error in packets when the VRVP scheme changes its rate.

To this end, we should fit the equation (11) properly in order to obtain the performance of adaptive modulation over the cognitive radio. Particularly, the switching thresholds are obtained as $\gamma_j = \gamma_K \cdot M$ with $M_j = j = 0,1,...,N$ according to the number of $N+1$ regions. These are actually the switching thresholds in case of VRVP scheme [7]. As mentioned above, the $\gamma_K$ optimization values are obtained by equation (8). In order to make our work more comprehensive, we give an instance of both switching thresholds in Table 1. With $\bar{\gamma} = 20$ is denoted the fading level in case of VRVP only, while with $N_t = 3$ is denoted the third transmission in case of adaptive modulation cross-layer combined with ARQ scheme. We have chosen to depict these values since they are the worst case of VRVP only and the better case of the cross-layer combination. Even in these cases, the switching thresholds of VRVP optimization are not high enough in order to avoid the full error in packets when the mode is changed.

*Table 1: The switching thresholds of VRVP vs. CLD*

|  | **No Transmit** | **BPSK** | **QPSK** | **16-QAM** | **64-QAM** |
|---|---|---|---|---|---|
| $\bar{\gamma} = 20$ | 0 | 5.2745 | 8.2848 | 14.3054 | 20.326 |
| $N_t = 3$ | 0 | 7.3765 | 10.4247 | 17.1872 | 23.1932 |

Figure 4 compare our results with those derived from the cross-layer design without considering cognitive radio channel allocation. In dashed red lines is depicted the results derived over cognitive radio and with blue lines are depicted the results derived from the conventional system presented in [9]. We illustrate the results for three retransmissions. This illustration shows the performance increase achieved when cognitive radio is assumed under cross-layer constraints. The performance gain is more outstanding in low average SNR regions. However, the lesson learned from this study is the way of making changes on fading levels when the cognitive radio is assumed. These changes are enforced by the spectrum pooling policy in which the cognitive radio is relied on. Henceforth, different changes should be taken place when other policies for sharing the spectrum sharing to each user. These changes are indicated by the channel gains at the physical layer which influence each scheme that involves such a physical layer i.e. with cognition capabilities.

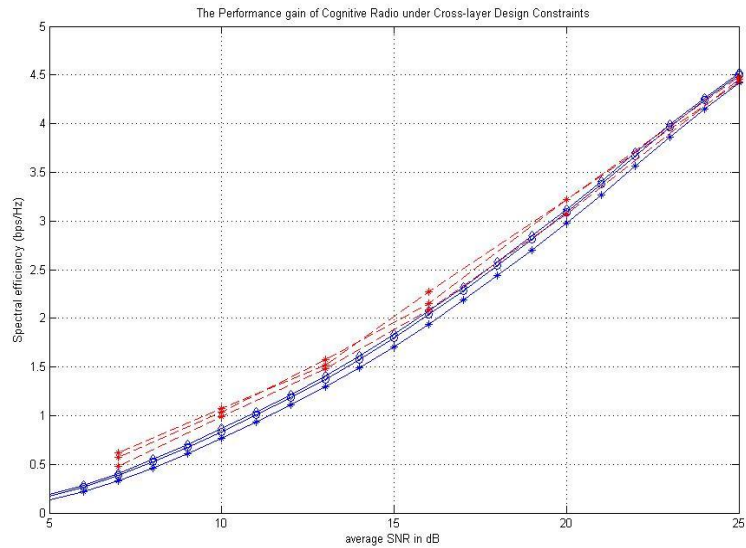

*Figure 4: The spectral efficiency of adaptive modulation over Cognitive Radio under cross-layer constraint*

## 6. Conclusions and Future work

This work presents the performance of physical layer over cognitive radio. At the first stage, we obtain the spectral efficiency of cognitive radio over fading channels. In sequel, we assume the adaptive modulation at the physical layer and we derive the spectral efficiency of such a scheme over radio with cognition capabilities. In addition to that, we study the performance of the above scheme under constraints imposed by a cross-layer combination of physical and data link layer. During this process, we realized the importance of spectrum pooling scheme of the cognitive radio. The spectrum pooling system indicates the channel gains at the physical layer that influences each scheme which involves such a physical layer. Finally, the numerical results of the performance analysis corroborate the performance gain of cognitive radio.

Next steps on this research topic could be the co-design of cognitive tasks and the several mechanisms implemented in different layers. For instance the spectrum sensing on the cognitive front-end should be designed considering several cross-layer combinations in order to accomplish the performance requirements of the upper layers. Besides, more opportunistic approaches, like the aforementioned spectrum pooling technique, should be assumed using cross-layer design criteria in order to assess the advantages of each approach across the layers. In conclusion, the deployment of cognitive radio into the wireless mobile networks available into the market should be studied using different design aspects.